\documentclass{epl}
\usepackage{amsmath,amssymb}
\usepackage{graphicx}
\title{Mean first-passage times for an ac-driven magnetic moment of a
nanoparticle}
\shorttitle{Mean first-passage times}
\author{S. I. Denisov\inst{1,2} \and  K. Sakmann\inst{1} \and
P. Talkner\inst{1} \and P. H\"{a}nggi\inst{1}}

\institute{
  \inst{1} Institut f\"{u}r Physik, Universit\"{a}t Augsburg -
  Universit\"{a}tsstra{\ss}e 1, D-86135 Augsburg, Germany \\
  \inst{2} Sumy State University - 2 Rimsky-Korsakov Street, 40007 Sumy,
  Ukraine}
\pacs{05.10.Gg}{Stochastic analysis methods} \pacs{75.60.Jk}{Magnetization
reversal mechanisms} \pacs{75.50.Tt}{Fine-particle systems; nanocrystalline
materials}

\begin{document}

\maketitle

\begin{abstract}
The two-dimensional backward Fokker-Planck equation is used to calculate the
mean first-passage times (MFPTs) of the magnetic moment of a nanoparticle
driven by a rotating magnetic field. It is shown that a magnetic field that is
rapidly rotating in the plane {\it perpendicular} to the easy axis of the
nanoparticle governs the MFPTs just in the same way as a static magnetic field
that is applied {\it along} the  easy axis. Within this framework, the features
of the magnetic relaxation and net magnetization of systems composed of
ferromagnetic nanoparticles arising from the action of the rotating field are
revealed.
\end{abstract}

\section{Introduction}
The mean first-passage time (MFPT), i.e., the average time that elapses until a
stochastic process reaches a prescribed domain, is an important characteristic
of the considered process. It is widely used for describing various dynamic
features such as exit problems, activation rates, lifetimes of metastable
states and other noise induced phenomena \cite{HTB,R}. The class of
stochastic processes for which the MFPT can be calculated explicitly is rather
limited \cite{HTB}. In fact, the most general analytical results were obtained
for continuous \textit{one-dimensional} Markov processes within an approach
based on the backward Fokker-Planck equation \cite{PAV,S,W2,GRD,Tal}. It is
important to note that most of the exactly known results are restricted to
Markov processes that are homogeneous in time.

However, Markov processes that describe \textit{time-depen\-dent} systems
usually cannot be approximated by homogeneous processes. Well-known
examples are, the phenomenon of Stochastic Resonance
\cite{GHJM,PH,RUBI} and directed transport in ratchet-like systems \cite{RH}
where external time-dependent fields play a crucial role. Since the
above-mentioned method is no longer applicable, the development of new
approaches for calculating MFPTs in periodically driven systems presents an
important challenge. For slowly varying external driving forces
adiabatic and semiadiabatic approximations are available \cite{TL,SKTH}. The
path integral formulation of the conditional probability provides a convenient
basis for approximations if the noise is weak \cite{Lehm}.

In this Letter we present an analytical approach to the
\textit{two-dimensional} MFPT problem for a rapidly driven magnetic moment of a
ferromagnetic nanoparticle. It is based on the backward Fokker-Planck equation
in a \textit{rotating} coordinate system, which describes the
\textit{homogeneous} dynamics of the magnetic moment, and on an averaging
procedure \cite{RTH}, which is similar to that used for describing the Kapitsa
pendulum with oscillating point of attachment \cite{K}. The analytical
results favorably compare with numerical simulations of the
equivalent Langevin equation.

\section{Basic equations}
In spherical coordinates, the Lan\-dau-Lifshitz equation \cite{LL2} for the
magnetic moment $\mathbf{m}(t) = m(\sin\theta \cos\varphi, \sin\theta
\sin\varphi, \cos\theta)$ of a single-domain ferromagnetic nanoparticle assumes the
form
\begin{equation}
    \dot{\theta} = -\displaystyle\frac{\gamma}{m\sin\theta}\left(\lambda
    \sin\theta\frac{\partial}{\partial\theta} + \frac{\partial}
    {\partial\varphi}\right)W, \quad 
    \dot{\varphi} = \displaystyle\frac{\gamma}{m\sin^{2}\theta}\left(\sin
    \theta\frac{\partial}{\partial\theta} - \lambda\frac{\partial}
    {\partial\varphi}\right)W,
    \label{eq:det_L-L}
\end{equation}
where $\theta = \theta(t)$ and $\varphi = \varphi(t)$ are the polar and
azimuthal angles of $\mathbf{m} (t)$, respectively, $\gamma(>0)$ is the
gyromagnetic ratio, $\lambda(>0)$ is a dimensionless damping parameter, $m =
|\mathbf{m} (t)|$ denotes the conserved total magnetic moment, and $W =
W(\theta, \varphi,t)$ is the magnetic energy of the nanoparticle. Besides the
damping, the interaction with a heat bath also generates stochastic elements in
the dynamics of the magnetic moment. These effects lead to a random
contribution to the effective magnetic field which is assumed to be given by
isotropic, Gaussian and white noise \cite{B,KH}.
The conditional probability density
$P(\theta,\varphi,t |\theta', \varphi',t')$ of the resulting Markovian
process satisfies the forward and the
backward Fokker-Planck equations which are equivalent to each other \cite{HT2}.
At equal times $P(\theta, \varphi, t'|\theta', \varphi',t') = \delta (\theta -
\theta') \delta (\varphi - \varphi')$, where $\delta(x)$ denotes the Dirac
$\delta$ function. Here we will use the backward equation, which propagates
$t'$ from $t$ on backward in time. In the present case it takes the form
\cite{DY}
\begin{equation}
    -\frac{1}{\Delta \gamma^2} \frac{\partial
    P(\theta,\varphi,t|\theta',\varphi',t')}{\partial t'} = L^+(t')
    P(\theta,\varphi,t|\theta',\varphi',t'),
    \label{bwf}
\end{equation}
where $\Delta = \lambda k_B T/\gamma m$ denotes the thermal noise intensity,
$k_B$ the Boltzmann constant, $T$ the temperature of the heat bath, $L^+(t')$
the backward Fokker-Planck operator,
\begin{equation}
    L^+(t')= \frac{\partial^{2}}{\partial\theta'^{2}} + \frac{1}
    {\sin^{2}\theta'}\frac{\partial^{2}}{\partial\varphi'^{2}} +
    (\cot\theta' + f(\theta',\varphi',t'))\frac{\partial }
    {\partial\theta'} + g(\theta',\varphi',t')\frac{\partial }
    {\partial\varphi'}, \nonumber
    \label{eq:bw_F-P}
\end{equation}
and the functions
\begin{equation}
\begin{array}{c}
    f(\theta,\varphi,t) = -\displaystyle\frac{1}{\Delta
    \gamma m \sin \theta} \left ( \lambda \sin \theta
    \frac{\partial}{\partial \theta} + \frac{\partial}{\partial
    \varphi} \right)W, \\[12pt]
    g(\theta,\varphi,t) = \displaystyle\frac{1}{\Delta \gamma m \sin^2
    \theta} \left ( \sin \theta \frac{\partial}{\partial \theta} -
    \lambda \frac{\partial}{\partial \varphi} \right)W
\end{array}
\label{fg}
\end{equation}
are proportional to the drift components of the angles $\theta$ and $\varphi$,
respectively, see eq.~(\ref{eq:det_L-L}). Note that if the energy $W$ depends
on time, the backward operator is also time dependent.

For the calculation of first-passage times a domain $\Omega(t)$ of the
magnetization state space, which is a sphere with radius $|m|$, has to be
specified. In general this domain may change upon evolving time. Starting out in
$\Omega(t')$ the magnetization eventually will leave this domain. The instant
$t_f$ of first crossing the boundary $\partial \Omega(t)$ defines the
first-passage time $t_f-t'$. In order to prevent the trajectory to recross this
boundary, absorbing boundary conditions must be imposed at $\partial
\Omega(t)$. For the backward equation they read
\begin{equation}
    P(\theta,\varphi,t|\theta',\varphi',t') =0 \quad \text{for}
    \;(\theta',\varphi') \in \partial \Omega(t').
    \label{bc}
    \end{equation}
The probability $P_{\Omega}(t|\theta',\varphi',t')$ that the magnetization has
been in the prescribed domain for all times, starting out at $t'$ up to time $t$, i.e.,
$\mathbf{m}(s) \in \Omega(s)$ for all $t' \leq s \leq t$, follows from the
conditional probability in the standard way:
\begin{equation}
    P_{\Omega}(t|\theta',\varphi',t') =\int_{\Omega(t)} d\theta\:d \varphi
    P(\theta,\varphi ,t|\theta',\varphi',t').
    \label{PO}
\end{equation}
The integral of $P_{\Omega}(t|\theta',\varphi',t')$ over all times $t$
determines the MFPT $T_\Omega =\langle t_f - t' \rangle$,
\begin{equation}
    T_\Omega(\theta',\varphi',t') =
    \int_{t'}^\infty dt P_{\Omega}(t|\theta',\varphi',t').
    \label{TO}
\end{equation}
Using all, the definition (\ref{TO}), the condition that $P_{\Omega} (t'|\theta',
\varphi', t') = 1$, and the backward Fokker-Planck equation (\ref{bwf}), we find
that the MFPT satisfies the following equation:
\begin{equation}
    \Delta \gamma^2 L^+(t') T_\Omega(\theta',\varphi',t') +
    \frac{\partial}{\partial t'} T_\Omega(\theta',\varphi',t') = -1.
    \label{mfpt}
\end{equation}
Accordingly, one obtains as boundary condition from eq.~(\ref{bc})
\begin{equation}
    T_\Omega(\theta',\varphi',t') = 0 \quad \text{for} \;(\theta',\varphi')
    \in \partial \Omega(t').
    \label{bcTO}
\end{equation}
For a periodic time dependence (with period $\mathcal{T}$) of the backward
operator and the $\Omega$-domain, i.e., for $L^+(t+\mathcal{T})= L^+(t)$ and
$\Omega(t+\mathcal{T})= \Omega(t)$, the asymptotic solution of
eq.~(\ref{mfpt}) is also periodic in time with the same period $\mathcal{T}$.
For time homogeneous processes and domains it is constant with respect to time
and fulfills the well known Pontryagin equation $\Delta \gamma^2L^+ T_\Omega =
-1$.

\section{General results}
We now consider the magnetic moment of a nanoparticle with uniaxial anisotropy
and assume that a static magnetic field $\mathbf{H}$ is applied along the easy
axis of magnetization which we choose as $z$ axis, i.e., $\mathbf{H} =
(0,0,H)$. Additionally, a rotating magnetic field $\mathbf{h}(t)$ acts
perpendicular to this axis, i.e., $\mathbf{h}(t) = h(\cos \omega t,\rho\sin
\omega t,0)$, where $\omega$ is the angular frequency and $\rho = -1,+1$
corresponds to clockwise and counterclockwise rotation, respectively. Hence,
the magnetic energy of the nanoparticle is given by
\begin{equation}
    W = \textstyle \frac{1}{2}mH_{a}\sin^{2}\theta - mH\cos\theta -
    mh \sin\theta \cos\psi
    \label{eq:W}
\end{equation}
with $H_{a}$ denoting the anisotropy field and $\psi = \varphi - \rho\omega t$.
We note that in this case a detailed analysis of the deterministic dynamics of
$\mathbf{m}(t)$ is presented in \cite{BSM}.
According to eq.~(\ref{eq:W}), the functions $f$ and $g$ in
eq.~(\ref{eq:bw_F-P}) depend on $\varphi'$ and $t'$ only through the single
variable $\psi' = \varphi' - \rho\omega t'$:
\begin{equation}
    \begin{array}{c}
    f =\! -2a(\cos\theta' \!+ \tilde{H})\sin\theta' \!+ \displaystyle
    \frac{2a\tilde{h}} {\lambda}(\lambda\cos\theta'\cos\psi' \!-
    \sin\psi'),  \\[10pt]
    g = \displaystyle\frac{2a}{\lambda}(\cos\theta' + \tilde{H}) -
    \frac{2a\tilde{h}}{\lambda\sin\theta'}(\lambda\sin\psi' +
    \cos\theta'\cos\psi'),  \\
    \end{array}
    \label{eq:f_g}
\end{equation}
where $a = mH_{a}/2k_{B}T$ is the anisotropy barrier height in units of
thermal energy $k_{B}T$, $\tilde{H} = H/H_{a}$, and $\tilde{h} =
h/H_{a}$.
By introducing a rotating frame, in which $\mathbf{h}(t)= h(1,0,0)$ and
the azimuthal angle $\varphi$ is replaced by $\psi$,
the time derivative $\partial /\partial t'$ goes over into
$\partial /\partial t' - \rho \omega \partial / \partial \psi'$, and
consequently the backward equation for the MFPT reads
\begin{equation}
    \frac{\partial^{2}T_{\Omega}}{\partial\theta'^{2}} +
    \frac{1}{\sin^{2}\theta'}\frac{\partial^{2}T_{\Omega}}{\partial
    \psi'^{2}} + (\cot\theta' + f)\frac{\partial T_{\Omega}}{\partial
    \theta'} + (g - \rho at_{r}\omega)\frac{\partial T_{\Omega}}
    {\partial\psi'} = -at_{r}\left (1+
      \frac{\partial T_\Omega}{\partial t'} \right ),
    \label{TOtd}
\end{equation}
where $t_r = 2/(\lambda \gamma H_a)$ is the characteristic relaxation time of
the precessional motion of the magnetic moment. Moreover, we assume that the
domain $\Omega(t)$ is stationary in the rotating frame, i.e., $\Omega(t)$ is
bounded by a curve $\phi_{\Omega} (\psi)$ on the sphere. Then, the steady-state
solution of (\ref{TOtd}) is independent of time and can be found as the
solution of the stationary Pontryagin equation
\begin{equation}
    \frac{\partial^{2}T_{\Omega}}{\partial\theta'^{2}} +
    \frac{1}{\sin^{2}\theta'}\frac{\partial^{2}T_{\Omega}}{\partial
    \psi'^{2}} + (\cot\theta' + f)\frac{\partial T_{\Omega}}{\partial
    \theta'} + (g - \rho at_{r}\omega)\frac{\partial T_{\Omega}}
    {\partial\psi'} = -at_{r}.
    \label{TOs}
\end{equation}
This presents already a major simplification of the original problem. However,
still a partial differential equation in two spatial dimensions remains to be
solved.

In the following we will consider two types of domains which we distinguish by
the index $\Omega := \pm 1$. The domain  with $\Omega =+1$ (denoted as
up domain) contains the up magnetization, $\theta =0$, and is bounded
in the rotating frame by a curve $\phi_{+1}(\psi)$. Accordingly, the
$\Omega=-1$ domain (down domain) contains the down magnetization,
$\theta = \pi$, and is bounded by a curve $\phi_{-1}(\psi)$. It is convenient
to present the respective MFPTs out of these domains in the form $T_{\Omega} =
\overline{T} _{\Omega}(\theta') + S_ {\Omega} (\theta', \psi')$, where the
overbar denotes an average over $\psi'$, i.e. $\overline {(\cdot)} = (1/2\pi)
\int_{0}^{2\pi} d\psi'(\cdot)$. Substituting the expansions $f = \overline{f} +
f_{1}$ and $g = \overline{g} + g_{1}$ into eq.~(\ref{TOs}), one obtains an
equation for $\overline{T}_{\Omega}$, reading
\begin{equation}
    \frac{d^{2}\overline{T}_{\Omega}}{d\theta'^{2}} + (\cot\theta'
    + \overline{f})\frac{d\overline{T}_{\Omega}}{d\theta'}
    + \overline{f_{1} \frac{\partial S_{\Omega}}{\partial\theta'}} +
    \overline{g_{1} \frac{\partial S_{\Omega}}{\partial\psi'}} =
    -at_{r}
    \label{eq:eq_mean_T}
\end{equation}
and an equation for $S_{\Omega}$, reading
\begin{equation}
    \frac{\partial^{2}S_{\Omega}}{\partial\theta'^{2}} +
    \frac{1}{\sin^{2}\theta'}\frac{\partial^{2}S_{\Omega}}{\partial
    \psi'^{2}} + (\cot\theta' + f)\frac{\partial S_{\Omega}}{\partial
    \theta'} + f_{1}\frac{d\overline{T}_{\Omega}}{d\theta'}
    + (g - \rho at_{r}\omega)\frac{\partial S_{\Omega}}
    {\partial\psi'} - \overline{f_{1} \frac{\partial S_{\Omega}}
    {\partial\theta'}} -  \overline{g_{1} \frac{\partial S_{\Omega}}
    {\partial\psi'}}= 0.
    \label{eq:eq_S}
\end{equation}
We emphasize that these equations are exact, i.e., they follow from the
backward Fokker-Planck equation (\ref{eq:bw_F-P}) for stationary domains in the
rotating frame.

\section{High-frequency limit}
In the case of an arbitrary rotating field we are not able to solve
eq.~(\ref{TOs}) analytically. But in the asymptotic limit of a fast rotating
field, i.e. for  $\omega \gg \omega_{r} \equiv \gamma H_{a}$,
eqs.~(\ref{eq:eq_mean_T}) and (\ref{eq:eq_S}) can be solved readily. The key
observation leading to their solution is that $S_{\Omega} \to 0$ for $\omega
\to \infty$. Assuming also that the derivatives of $S_\Omega$ tend to zero as
$1/\omega$ for $\omega \to \infty$, eq.~(\ref{eq:eq_S}) simplifies to read in
this high-frequency limit
\begin{equation}
    \rho\,a\,t_{r}\,\omega\frac{\partial S_{\Omega}}{\partial\psi'} -
    f_{1}\frac{d\overline{T}_{\Omega}}{d\theta'} = 0.
    \label{eq:eq_S1}
\end{equation}
With $f_{1} = (2a\tilde{h}/\lambda) (\lambda \cos\theta' \cos\psi' -
\sin\psi')$, see eq.~(\ref{eq:f_g}), the solution of this equation that
satisfies the condition $\overline{S}_{\Omega} = 0$ can be written in the form
\begin{equation}
    S_{\Omega} = \rho \frac{\tilde{h}}{\tilde{\omega}}(\lambda\cos\theta'
    \sin\psi' + \cos\psi')\frac{d \overline{T}_{\Omega}}{d\theta'},
    \label{eq:S1}
\end{equation}
where $\tilde{\omega} = \omega/\omega_{r}$. 
Note, that according to the above assumptions, this solution is valid if
$\tilde{h}/\tilde{\omega} \ll 1$. The $\omega$ dependence assumed in the
derivation of eq. (\ref{eq:eq_S1}) is now confirmed selfconsistently. Using
eq.~(\ref{eq:S1}) and the relation $g_{1} = -(2a\tilde{h} /\lambda)
(\lambda\sin\psi' + \cos\theta' \cos\psi') /\sin\theta'$, we proceed to
calculate the averages
\begin{equation}
    \overline{f_{1} \frac{\partial S_{\Omega}}{\partial\theta'}} =
    \overline{g_{1} \frac{\partial S_{\Omega}}{\partial\psi'}} =
    -a\tilde{h}_{\text{eff}}\sin\theta'\frac{d \overline{T}_{\Omega}}{d\theta'}
    \label{eq:means}
\end{equation}
with $\tilde{h}_{\text{eff}} = -\rho\tilde{h}^{2}/\tilde{\omega}$. Finally,
substituting eq.~(\ref{eq:means}) and $\overline{f} = -2a(\cos\theta' +
\tilde{H}) \sin \theta'$ into eq.~(\ref{eq:eq_mean_T}), we obtain the desired
equation for $\overline{T}_{\Omega}$ in the high-frequency limit:
\begin{equation}
    \frac{d^{2}\overline{T}_{\Omega}}{d\theta'^{2}} + [\cot\theta'
    - 2a(\cos\theta' + \tilde{H} + \tilde{h}_{\text{eff}})\sin\theta']
    \frac{d\overline{T}_{\Omega}}{d\theta'} = -at_{r}. \\
    \label{eq_mean_T2}
\end{equation}

This equation exhibits the remarkable result that a magnetic field that is
rapidly rotating in the plane \textit{perpendicular} to the easy axis of the
nanoparticle acts on the nanoparticle's magnetic moment precisely as a static
magnetic field $\tilde{h}_{\text{eff}}$ (in units of $H_{a}$) which is applied
\textit{along} the easy axis. The direction of the effective field
$\tilde{h}_{\text{eff}}$ and the direction of the field rotation follow the
left-hand rule, and the value of $\tilde{h}_{\text{eff}}$ is the same for up
and down domains. It is important to emphasize that although the condition
$\tilde{h}/ \tilde{\omega} \ll 1$ holds, the effective field can be large if
$\tilde{h} \gg 1$.

We note that $\theta'=0,\pi$ are singular points of
eq.~(\ref{eq_mean_T2}) for $\Omega =+1,\,-1$, respectively. At these points the
general solution of eq.~(\ref{eq_mean_T2}) exhibits logarithmic singularities.
To prevent this non-physical behavior of $\overline{T} _\Omega$ the
regularity condition $d\overline{T}_\Omega / d\theta' |_{\theta' = \pi(1 -
\Omega)/2} = 0$ must hold \cite{DLT}. In order to derive the boundary condition
for eq.~(\ref{eq_mean_T2}), we decompose the function $\phi _{\Omega} (\psi)$
into its average and its periodic parts, i.e., $\phi_ \Omega (\psi) = \overline
{\phi}_\Omega + \vartheta_{\Omega}(\psi)$. Assuming that $|\vartheta
_{\Omega}(\psi)| \ll \overline{\phi}_\Omega$, from the absorbing boundary
condition of the full two-dimensional problem, $T_{\Omega} (\phi _{\Omega}
(\psi'),\psi') = 0$, we find the periodic part of $\phi_ \Omega (\psi)$,
$\vartheta_{\Omega}(\psi) = - \rho\tilde{h} (\lambda \cos \overline {\phi}
_\Omega \sin \psi + \cos \psi) / \tilde{\omega}$, and the desired boundary
condition for the averaged MFPT, $\overline{T} _\Omega (\overline {\phi}
_\Omega) = 0$. The solution of eq.~(\ref{eq_mean_T2}) with the specified
regularity and boundary conditions becomes
\begin{equation}
    \overline{T}_{\Omega}(\theta') = at_{r}\int_{\cos\overline{\phi}_
    {\Omega}}^{\cos \theta'}dx \frac{e^{-a(x + \tilde{H} + \tilde{h}_
    {\text{eff}})^{2}}}{1 - x^{2}} \int_{x}^{\Omega}dy\, e^{a(y +
    \tilde{H} + \tilde{h}_{\text{eff}})^{2}},
    \label{eq:mfpt}
\end{equation}
where $\theta' \in [0,\overline{\phi}_{+1}]$ if $\Omega = +1$, $\theta' \in
[\overline{\phi}_{-1},\pi]$ if $\Omega = -1$, and the angles $\overline
{\phi}_{\Omega}$ can be chosen depending on physical situation.

In the case of a high potential barrier, $a \gg 1$, and moderately
large total effective fields, $|\tilde{H} + \tilde{h}_{\text{eff}} |
< 1$, the magnetic moment resides near one of two equilibrium
directions, up or down. Since transition times between these states
by far exceed the relaxation times towards these states, the
averaged MFPT $\overline{T} _{\Omega}(\theta')$ describing the
transition from one state $\Omega$ to the opposite
state $-\Omega$ only weakly depends on the precise location of the
initial magnetization, as long as $\theta'$ lies within the domain
of attraction of the considered state $\Omega$. Also the precise
location of the absorbing boundary $\overline {\phi}_ {\Omega}$ has
practically no influence on $\overline{T} _{\Omega} (\theta')$ if it
is located well beyond the separatrix which divides the state space
into domains of attraction of the up and down magnetization.
Then the eq.~(\ref{eq:mfpt}) yields in leading order in $a$
\begin{equation}
    \overline{T}_{\Omega} = t_{r}\sqrt{\frac{\pi}{a}}\,\frac{e^{a[1 +
    \Omega(\tilde{H}+\tilde{h}_{\text{eff}})]^2}}{2[1 -
    (\tilde{H}+\tilde{h}_{\text{eff}})^2][1 +
    \Omega(\tilde{H}+\tilde{h}_{\text{eff}})]}.
    \label{eq:mfpt2}
\end{equation}
Using eqs.~(\ref{eq:S1}), (\ref{eq:mfpt}) and (\ref{eq:mfpt2}), it is not too
difficult to demonstrate  that $|S_{\Omega}| \ll |\Delta \overline{T}
_{\Omega}|$, where $\Delta \overline{T}_{\Omega} = \overline{T}_{\Omega} -
\overline{T}_{\Omega} |_{\tilde{h} = 0}$ is the contribution of $\tilde{h}$ to
$\overline{T} _{\Omega}$. This means that the periodic part of $T_{\Omega}
(\theta',\psi')$ can be neglected such that $T_{\Omega} (\theta',\psi') \approx
\overline{T} _{\Omega}$.

\begin{figure}[htpb]
  \begin{center}
    \includegraphics[angle=0,width=0.5\linewidth]{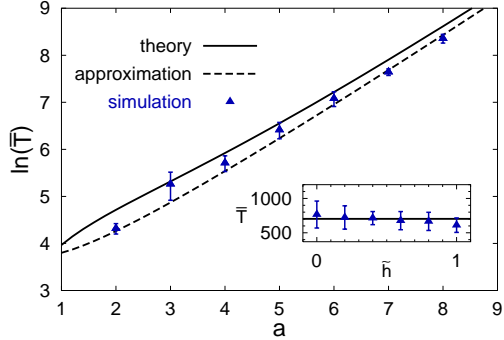}
  \end{center}
  \caption{(Color online) The natural logarithm of the dimensionless
    MFPT, $\overline{T} =\overline{T}_{+1}\cdot\omega_{r}$, is
    displayed as a function of  the dimensionless anisotropy barrier height
    $a$. The solid line
    is the result of present theory (\ref{eq:mfpt}), and the broken
    line depicts the approximate relation (\ref{eq:mfpt2}). The
    symbols indicate results from the numerical simulation of $4
    \times 10^4$ runs of the Langevin equation which is equivalent to the
    backward equation (\ref{bwf}) with absorbing boundary at
    $\theta'_{\text{abs}}=0.8 \pi$ and a starting point at
    $\theta'=0.05\pi$, $\phi'=0$.
    The values of the other parameters
    are $\lambda=0.1$, $\rho =1$, $\omega = 10\: \omega_r$, $\tilde{H}=0$
    and $\tilde{h}_{\text{eff}} = -0.1$.  $\bar{T}$ is
    displayed in the inset as a function of $\tilde{H}$, where  $\tilde{H}
    +\tilde{h}_{\text{eff}} = -0.1$, $a=5$, $\omega=10\,\omega_r$,
    $\lambda=0.1$ and $\rho=1$. Only small systematic deviations from
    the theoretical prediction (solid line) are visible.
    }
  \label{fig1}
\end{figure}

In Fig.~\ref{fig1} the theoretical prediction (\ref{eq:mfpt}) and
the asymptotic approximation (\ref{eq:mfpt2}) are compared with
the results of a numerical simulation of the coupled Langevin
equations for the two angles $\theta$ and $\phi$ which are equivalent
to the process described by the eq.~ (\ref{bwf}).

Next we examine the most interesting case of zero static field, $\tilde{H} =
0$, when only the effective field influences the MFPTs. Specifically, if
$|\tilde{h}_{\text{eff}}| \ll 1$ then eq.~(\ref{eq:mfpt2}) yields $\overline{T}
_{\Omega} = \tau_{0} \exp(\Omega2 a\tilde{h} _{\text{eff}})$, where $\tau_{0} =
(t_{r}/4) \sqrt{\pi/a} \exp a$. Thus, the rotating magnetic field increases the
MFPT for the magnetic moment $\mathbf{m} (t)$ in the state $\Omega =
-\rho$ and decreases it for $\mathbf{m}(t)$ in the state $\Omega = \rho$,
where $\rho = \pm 1$. Physically, this difference in MFPTs
follows from the natural counter-clockwise precession (if looked from above)
of the magnetic moment.
Therefore, for the two directions of the magnetic field
rotation, $\rho = -1$ and $\rho = +1$, the forced dynamics of $\mathbf{m}(t)$
in the up and down states is different. Hence, the rotating magnetic
field breaks the degeneracy between the up and down orientations of the
magnetic moment for $\tilde{H} = 0$.

\section{Magnetic relaxation}
To illustrate the role of the effective magnetic field $\tilde{h}_
{\text{eff}}$, we consider the relaxation of magnetization in a system composed
of ferromagnetic nanoparticles whose easy axes are perpendicular to the plane
of field rotation and $a \gg 1$. The reduced magnetization of this system can
be defined as $\mu(t) = [N_{+1}(t) -N_{-1}(t)]/N$, where $N(\gg 1)$ and
$N_{\Omega}(t)$ denote the total number of nanoparticles and those that are in
the state $\Omega$, respectively. Using $N_{-1}(t) + N_{+1} (t) = N$, we obtain
$\dot{\mu}(t) = 2\dot{N}_{+1}(t)/N$.  On the other hand, the time dependence of
$N_\Omega(t)$ is governed by the kinetic equation $\dot{N}_ {\Omega} (t) =
N_{-\Omega}(t) w_{-\Omega} - N_{\Omega}(t) w_{\Omega}$, where $w_\Omega$
denotes the transition rate from $\Omega$ to $-\Omega$. This gives for the
magnetization the well-known equation
\begin{equation}
    \dot{\mu}(t) = -\mu(t)(w_{-1} + w_{+1}) - w_{+1} + w_{-1}.
    \label{eq:eq_relax}
\end{equation}

Because the mean residence time in the state $\Omega$ equals $\overline{T}
_{\Omega}$, the transition rate $w_{\Omega}$ is given by $w_{\Omega} = 1/
\overline{T}_{\Omega}$. In this case, solving eq. (\ref{eq:eq_relax}) with the
initial condition $\mu(0) = 1$ and assuming $\tilde{H} = 0$, we obtain the
relaxation law
\begin{equation}
    \mu(t) = (1 -\mu_{\infty})\exp[-t/\tau] + \mu_{\infty},
    \label{eq:relax_law}
\end{equation}
where $\tau = 1/(w_{-1} + w_{+1}) = \tau_{0}/\cosh(2a\tilde{h}_{\text{eff}})$
and $\mu_{\infty} = (w_{-1} - w_{+1})/(w_{-1} + w_{+1}) = \tanh(2a \tilde{h}
_{\text{eff}})$ are the relaxation time and steady-state magnetization,
respectively. Thus, the rotating magnetic field decreases the relaxation time
and magnetizes the nanoparticle system. These results are not evident because
they arise from the difference between the up and down dynamical states of the
magnetic moments. It is important to note in this context that even for small
values of the effective field the magnetization effect can be sizable. In
particular, if $\tilde{h} = 0.1$ and $\tilde{\omega} = 10$ then $|\tilde{h}
_{\text{eff}}| = 10^{-3}$, and for $a = 50$ we get $|\mu_{\infty}| = 0.1$.

\section{Conclusion}
We have shown that a magnetic field rapidly rotating in the plane perpendicular
to the easy axis of a nanoparticle, lifts the degeneracy between the up and
down orientations of the nanoparticle magnetic moment. This lifting is
characterized by the effective magnetic field acting along the easy axis in a
direction that is uniquely defined by the direction of the magnetic field
rotation. The effective field changes the MFPTs for the nanoparticle's magnetic
moment and, as a consequence, changes the relaxation law and yields a net
magnetization for a system of ferromagnetic nanoparticles.

\acknowledgments S.I.D. acknowledges the support of the EU through
NANOSPIN project, contract No NMP4-CT-2004-013545, and through a
Marie Curie individual fellowship, contract No MIF1-CT-2005-007021,
P.T. and P.H. acknowledge the support of the DFG via the SFB 486.

\end{document}